\documentclass[prd,aps,preprintnumbers,nofootinbib,twocolumn]{revtex4}
\usepackage{epsfig}
\usepackage{amsmath}
\usepackage{hyperref}

\begin{document}

\preprint{CERN-PH-TH/2011-026}

\renewcommand{\thefigure}{\arabic{figure}}

\title{The like-sign dimuon charge asymmetry at the Tevatron:\\ corrections from $B$ meson fragmentation}

\author{Alexander Mitov}
\affiliation{Theory Division, CERN, CH-1211 Geneva 23, Switzerland}

\date{\today}

\begin{abstract}
The existing predictions for the like-sign dimuon charge asymmetry at the Tevatron are expressed in terms of parameters related to $B$ mesons' mixing and inclusive production fractions. We show that in the realistic case when phase-space cuts are applied, the asymmetry depends also on the details of the production mechanism for the $B$ mesons. In particular, it is sensitive to the difference in the fragmentation functions of $B^0_d$ and $B^0_s$ mesons. We estimate these fragmentation effects and find that they shift the theory prediction for this observable by approximately 10\%. We also point out the approximately 20 \% sensitivity of the asymmetry depending on which set of values for the $B$ meson production fractions is used: as measured at the Z pole or at the Tevatron. The impact of these effects on the extraction of $A^s_{\rm SL}$ from the D\O\ measurement is presented.
\end{abstract}
\maketitle

\section{Introduction}

Among the most natural places to look for deviations from the predictions of the Standard Model (SM) are CP-violation sensitive observables, since CP-violation in the SM is small. In a recent publication \cite{Abazov:2010hv}, the D$\O$ collaboration presented evidence for a $3.2\sigma$ deviation from the SM. Together with the discrepancy in top-quark forward-backward asymmetry \cite{Aaltonen:2011kc}, this is the most significant current deviation from the SM. The implication of the measurement \cite{Abazov:2010hv} is that the discrepancy might be due to a beyond the SM CP-violating effect in the neutral $B$ meson system. 

Given the intense interest in the results of Ref.~\cite{Abazov:2010hv} and their potential implication for beyond the SM  (bSM) physics, in sec.~\ref{sec:THanalysis} we revisit the analysis with emphasis on $B$ meson production effects characteristic for hadron colliders. To the best of our knowledge such effects have not been consider previously. In sec.~\ref{sec:put-together} we estimate the numerical impact of various higher order perturbative and non-perturbative effects on the reported discrepancy. 

Accounting for $B$-production effects, in sec.~\ref{sec:discussion} we consider the extraction of the flavor specific asymmetry $A^s_{\rm SL}$. We observe that the inclusion of these effects further strengthens the expectation that {\it both} $A^s_{\rm SL}$ and $A^d_{\rm SL}$  are deviating from their SM values.

\section{Overview of the measurement}\label{sec:Overview}

The D$\O$ collaboration measures \cite{Abazov:2010hv} the like-sign dimuon charge asymmetry $A_{\rm exp}$, defined as
\begin{equation}
A_{\rm exp}\equiv {N^{++}-N^{--} \over N^{++}+N^{--} } \, ,
\label{eq:A}
\end{equation} 
where $N^{++} (N^{--})$ represent the number of events containing at least two positively (resp. negatively) charged muons. $A_{\rm exp}$ contains both signal and backgrounds.

The signal is defined as the same charge muon pairs originating from semileptonic decays of intermediate neutral $B$ mesons:
\begin{eqnarray}
A^b_{sl} \equiv {N_b^{++}-N_b^{--} \over N_b^{++}+N_b^{--} } \, ,
\label{eq:Ab-sl}
\end{eqnarray} 
and is usually expressed as:
\begin{eqnarray}
A^b_{sl} = {N(B^0\bar{B}^0 \to \mu^+\mu^+ X)-N(B^0\bar{B}^0 \to \mu^-\mu^- X) 
\over N(B^0\bar{B}^0 \to \mu^+\mu^+ X)+N(B^0\bar{B}^0 \to \mu^-\mu^- X)  } \, .\nonumber
\end{eqnarray} 
As we will show in the following, the production mechanism for the intermediate $B\bar{B}$ pairs does not decouple and also contributes to $A^b_{sl}$.

In the context of Eq.~(\ref{eq:Ab-sl}), the origin of the asymmetry is due to an oscillation $B^0 \leftrightarrow \bar{B}^0$ followed by a decay to a muon of the ``wrong" sign \cite{Grossman:2006ce,Lenz:2006hd,Anikeev:2001rk}. Note that the intermediate neutral $B$ mesons in Eq.~(\ref{eq:Ab-sl}) are a mixture of $B^0_s$ and $B^0_d$ mesons. 

An assumption built into the experimental analysis \cite{Abazov:2010hv}, and one we also adopt here, is that at parton level the production stage is symmetric in $b$ and $\bar{b}$. We will comment on it in the next section.

The measurement involves the following cuts: $|\eta|<2.2$ and $p_T$ in the range $4.2 {\rm GeV} < p_T < 25 {\rm GeV}$. Additionally, muons with $p_T$ in the range $1.5 {\rm GeV} < p_T < 4.2 {\rm GeV}$ are included if they have longitudinal momentum component $|p_Z| > 6.4 {\rm GeV}$. The additional requirement $|p_Z| > 6.4{\rm GeV}$ for $p_T$'s in the range $1.5 {\rm GeV} < p_T < 4.2 {\rm GeV}$ 
is implemented by D\O\ in order to recover the acceptance for some of the muons in the forward region that would otherwise be lost due to the $p_T$ cuts.

There are two types of backgrounds contributing to Eq.~(\ref{eq:A}). The first type is mostly due to Kaons (see Ref.~\cite{Abazov:2010hv} for details); it contributes to Eq.~(\ref{eq:A}) but not to Eq.~(\ref{eq:Ab-sl}). The asymmetry in these events is due to a mechanism totally different from the one for the signal (essentially it is a detector effect). This type of background is roughly twice as large as the signal; it has to be subtracted from the data first. It is explained in \cite{Abazov:2010hv} that this background is known with high precision.

The second type of background consists of same sign muon pairs that contribute only to the denominators of Eqs.~(\ref{eq:A}) and (\ref{eq:Ab-sl}). The contribution from this type of background is relatively small \cite{Abazov:2010hv}; see appendix \ref{sec:app}.

\section{Theoretical analysis}\label{sec:THanalysis}

The next step is to derive the SM prediction $A_{\rm th}$ for the quantity $A^b_{sl}$ as defined in Eq.~(\ref{eq:Ab-sl}). In doing this we ignore the charge asymmetry $\Delta$ related to detection and identification of muons (see Ref.~\cite{Abazov:2010hv}) and assume it has been dealt with at the level of the data. We also ignore all backgrounds (discussed in appendix \ref{sec:app}). This way we consider only ``signal" events that have un-equal probabilities for producing same-sign muon pairs of opposite charge.

There are two such mechanisms. The first one is the signal described after Eq.~(\ref{eq:Ab-sl}) where the same-sign muon pairs are resulting from the decay of intermediate $B$ mesons {\it after} an oscillation. An example is the reaction $b\to \bar{B} \to \mu^-~;~ \bar{b}\to B\to \bar{B}\to \mu^-$. The asymmetry in this reaction is due to CP-violation in the neutral $B$ meson system \cite{Grossman:2006ce,Lenz:2006hd,Anikeev:2001rk} and is the subject of this study. 

As a second possibility, same-sign dimuon asymmetry can be generated from asymmetry in the production of $b\bar{b}$ pairs. A $b\bar{b}$ asymmetry cannot be generated perturbatively within QCD, but at the non-perturbative level DGLAP evolution {\it predicts} non-zero asymmetry \cite{Catani:2004nc} for the proton's $b\bar{b}$ parton distributions. While at the Tevatron this should not produce any net asymmetry (since it is a $p\bar{p}$ machine) it might be of relevance for the LHC.

\subsection{Calculational framework}

The number of same-sign muon pairs $N(p{\bar p}\to B^0\bar{B}^0 \to \mu^a\mu^a X)$, with $a=\pm$, can be derived within the framework of (heavy) di-hadron production at hadron colliders, including the decay of the produced hadrons. One has to calculate the partonic cross-section $d\sigma(p\bar{p}\to b \bar{b} +X)$ differential in the rapidities and $p_T$ of the observed quarks, as appropriate, and with full account of $m_b$. Then one has to convolute it with non-perturbative fragmentation functions describing the long-distance quark-meson transition $b\to B_q$, $q=d,s$ for each one of the two $b,\bar{b}$ quarks, followed by the decay $B_q\to \mu+X$. For more details about $b$-production at hadron colliders see, for example, Refs.~\cite{Cacciari:2002pa,Cacciari:2003uh}. Adopting concise notations we have:
\begin{equation}
N^{\pm\pm}({\rm cuts}) = \int_{\rm cuts}~ d{\rm PS} ~ d\sigma^{\pm\pm} \, ,
\label{eq:N+-}
\end{equation}
with:
\begin{eqnarray}
d\sigma^{++} &=& \sum_{i,j=d,s} d\sigma^{\rm pert}_{b\bar{b}} \otimes_b \left( \hat{D}^{\rm np}_{b\to \bar{B}_i} \otimes \hat{D}^{\rm W}_{\bar{B}_i\to \mu^+} \right)_b \nonumber\\
&&~~~~~~~~~ ~~~~~~~ \otimes_{\bar{b}} \left( \hat{D}^{\rm np}_{\bar{b}\to {B}_j} \otimes \hat{D}^{\rm R}_{B_j\to \mu^+} \right)_{\bar{b}} \, , \label{eq:dsigma++}\\
&& \nonumber\\
d\sigma^{--} &=& \sum_{i,j=d,s} d\sigma^{\rm pert}_{b\bar{b}} \otimes_b \left( \hat{D}^{\rm np}_{b\to \bar{B}_i} \otimes \hat{D}^{\rm R}_{\bar{B}_i\to \mu^-} \right)_b \nonumber\\
&&~~~~~~~~~ ~~~~~~~ \otimes_{\bar{b}} \left( \hat{D}^{\rm np}_{\bar{b}\to {B}_j} \otimes \hat{D}^{\rm W}_{B_j\to \mu^-} \right)_{\bar{b}} \, .
\label{eq:dsigma--}
\end{eqnarray} 
In Eq.~(\ref{eq:N+-}) we have omitted an overall factor  that cancels in $A_{\rm th}$, and $\otimes_b \left(\dots\right)_b$ denotes a convolution with respect to the variables of the parton $b$; similarly for $\bar{b}$. 

The functions $\hat{D}^{\rm np}$ appearing in Eqs.~(\ref{eq:dsigma++},\ref{eq:dsigma--}) are the non-perturbative fragmentation functions for the $b\to B$ transition that are detailed in sec.~\ref{sec:fragment} and $\hat{D}^{\rm R,W}$ are the ``right" and ``wrong" $B$ meson decay functions detailed in sec.~\ref{sec:decay}.

The functions $d\sigma^{\rm pert}_{b\bar{b}}$ in Eqs.~(\ref{eq:dsigma++},\ref{eq:dsigma--}) are the corresponding partonic cross-sections for producing a $b\bar{b}$ pair. Dihadron production has been studied in the massless case \cite{Chiappetta:1996wp,Owens:2001rr,Binoth:2002wa,Almeida:2009jt}. It can also be applied to the massive case within the FONLL formalism of Ref.~\cite{Cacciari:1998it}, with the recent advances of Ref.~\cite{Biswas:2010sa} and with the good theoretical control over heavy flavor fragmentation \cite{Mele:1990cw,Melnikov:2004bm,Mitov:2004du,Mitov:2006ic}. 

The intrinsic hard scale in the process is the invariant mass of the {\it well-separated} quark pair,  which makes it possible to calculate observables even for small $p_T$
\footnote{Note that since the $b$-quarks are massive, $m_b>>\Lambda_{\rm QCD}$, the hard scale is bounded from below by $m_b$.}.

As we have emphasized in the notation adopted in Eq.~(\ref{eq:N+-}), the number of produced same-sign muon pairs depends on the cuts; the set of cuts is spelled out at the end of sec.~\ref{sec:Overview}. Moreover, in presence of cuts, the effects of quark production, fragmentation and decay do not decouple from each other - even after integration over the phase space. It is only in the limit of no cuts (i.e. unrestricted integration over the whole phase space) that such factorization might take place.

Important assumptions have been built into Eqs.~(\ref{eq:N+-},\ref{eq:dsigma++},\ref{eq:dsigma--}). First, we have summed over all intermediate states which we take as $B^0_s$ and $B^0_d$. This is a standard assumption in works on the subject. Second, we have assumed that the production of the intermediate neutral $B$ mesons factorizes from their decay. This is natural since the $B_{d,s}$-mesons are well defined particles and different scales drive their production/decay.

\subsection{$b\to B$ fragmentation functions}\label{sec:fragment}

The non-perturbative fragmentation functions $\hat{D}^{\rm np}_{b\to B}$ appearing in Eqs.~(\ref{eq:dsigma++},\ref{eq:dsigma--}) are typically extracted from LEP data. We normalize them:
\begin{equation}
\hat{D}^{\rm np}_{b\to \bar{B}_q} = f_q D^{\rm np}_{b\to \bar{B}_q} \, ,
\label{eq:Dhat-D}
\end{equation}
where $\int_0^1 dx D^{\rm np}_{b\to \bar{B}_q} = 1$, and $f_q$ are measured production fractions. We introduce the following simplified notation:
\begin{equation}
D^{\rm np}_{b\to \bar{B}_q} = D^{\rm np}_{\bar{b}\to B_q} \equiv D_{B_q} \, .
\label{eq:DbB-relation}
\end{equation}

The fractions $f_{s,d}$ are measured at the $Z$ pole and at the Tevatron. The two sets of measurements differ from each other by roughly $20\%$, although the fractions are generally assumed to be universal. The most up-to-date values can be found in Table 4 of Ref.~\cite{TheHeavyFlavorAveragingGroup:2010qj}. Indeed, as long as the factorization theorem applies, they have to be process independent since they refer to the non-perturbative transition $b\to B$ which takes place at scale $\sim m_b << Q$. In the analysis of Ref.~\cite{Abazov:2010hv} the values measured at the Tevatron are used. 

The fragmentation functions $D_{B_s}$ and $D_{B_d}$ are not known. The most accurately measured fragmentation functions at the $Z$ pole are for a combination of $B^\pm, B^0_d, B^0_s$ mesons and $\Lambda_b$ \cite{Heister:2001jg,Abe:2002iq,Abbiendi:2002vt}. We expect that these two functions are different. First, the two fragmentation fractions are rather different (by approximately a factor of 4) which implies different non-perturbative dynamics in the two cases. Second, we can {\it roughly} estimate these functions by utilizing the  fragmentation function of Peterson et al. \cite{Peterson:1982ak}:
\begin{equation}
D_{B_q}(x) = {\cal N} ~ {1\over x}\left( 1-{1\over x}-{\epsilon_q\over 1-x}\right)^{-2} \, ,
\label{eq:Peterson}
\end{equation}
where ${\cal N}$ is fixed by the normalization and we take the parameter $\epsilon_q=m_q^2/m_b^2$. In the following we will use $m_b=4.75 {\rm GeV}$ (as in Ref.~\cite{Cacciari:2002pa}) and $m_d=70 {\rm MeV}, m_s=170 {\rm MeV}$. Our motivation for choosing these values for $m_{d,s}$ is the following. The best value for $\epsilon$ extracted in Ref.~\cite{Cacciari:2002pa} is $\epsilon=0.002$, and it tends to decrease with the inclusion of higher order effects (see also \cite{Cacciari:2002re}). This parameter is for the above mentioned combination of $b$-flavored hadrons. Assuming that the fragmentation functions for the charged mesons and baryons are softer than these for the $B^0_{d,s}$, our choice of $m_{d,s}$ leads to the following reasonable values for the fragmentation parameters $\epsilon_s=0.0013,~\epsilon_d=0.0002$. A second constraint we have imposed on $m_{d,s}$ is that the difference $m_s-m_d$ is consistent with the mass difference of the corresponding $B$ mesons (or of the current quark masses). 

In our subsequent analysis we need the fifth moments $N=5$ of these functions $D_{B_q}^{(N)} = \int_0^1x^{N-1} D_{B_q}(x) dx$ for which we obtain $D_{B_s}^{(5)} = 0.69$ and $D_{B_d}^{(5)} = 0.82$. In fact, the only information about the functions $D_{B_q}$ that will enter our analysis is their ratio 
\begin{equation}
\kappa = D_{B_s}^{(5)}/D_{B_d}^{(5)} = 0.84  \, .
\label{eq:kappa}
\end{equation}

We stress once again that our method of estimation should be adequate for its restricted use. Furthermore our estimates are consistent with the expectation for the difference between these fragmentation function discussed in Ref.~\cite{Jaffe:1993ie}.

\subsection{$B\leftrightarrow \bar{B}$ oscillations and $B\to \mu+X$ decay spectra}\label{sec:decay}

There are two types of decays for the produced $B$ mesons: ``right" (R) decays $B\to \mu^+ + X$ (or $\bar{B}\to \mu^- +X$) and ``wrong" (W) ones $B\to \mu^- + X$ (or $\bar{B}\to \mu^+ +X$). For semileptonic decays the ``wrong" decays are due to an oscillation $B \leftrightarrow \bar{B}$ followed by a ``right" decay (see Refs.~\cite{Anikeev:2001rk} for more information). The ``right" decay is assumed to contain no direct CP violation \cite{Lenz:2006hd}. Also, oscillation and decay are assumed to decouple from each other. Thus, the decay functions in Eqs.~(\ref{eq:dsigma++},\ref{eq:dsigma--}) read:
\begin{eqnarray}
&&\hat{D}^{\rm R}_{B_q\to \mu^+} = \hat{D}^{\rm R}_{\bar{B}_q\to \mu^-} \equiv T(B_q\to B_q) \Gamma_q D^R_q \, ,\nonumber\\
&&\hat{D}^{\rm W}_{B_q\to \mu^-} \equiv T(B_q\to \bar{B}_q) \Gamma_q D^R_q \, ,\nonumber\\
&&\hat{D}^{\rm W}_{\bar{B}_q\to \mu^+} \equiv T(\bar{B}_q \to B_q) \Gamma_q D^R_q \, .
\label{eq:DR}
\end{eqnarray}

The decay function $D^R_q$ in Eq.~(\ref{eq:DR}) is normalized $\int_0^1 D^R_q(x) dx = 1$, and $\Gamma_q$ is the semileptonic decay width of $B_q$. We have also made use of the following result about the time-integrated probabilities $T(\bar{B}_q\to \bar{B}_q)=T(B_q\to B_q)$ \cite{Grossman:2006ce}. The semileptonic widths for $B^0_s$ and $B^0_d$ mesons are almost equal $\Gamma_s\approx \Gamma_d$. In the following we will consider that the decay functions are also equal, i.e. $D^R_s=D^R_d$. We will not need their explicit form.

\subsection{Putting everything together}\label{sec:put-together}

Combining the results above, we get the following result for the number of same-sign muon pairs:
\begin{eqnarray}
&& N^{\pm\pm}({\rm cuts}) = \sum_{i,j=s,d} N^{\pm\pm}_{ij}({\rm cuts}) \, ,
\label{eq:N+--simplified}\\
&& N^{++}_{ij} =  f_if_j \Gamma_i\Gamma_j T(B_i\to B_i) T(\bar{B}_j \to B_j) f_{ij}({\rm cuts})\, , 
\nonumber\\
&& N^{--}_{ij} =  f_if_j \Gamma_i\Gamma_j T(B_i\to B_i) T(B_j \to \bar{B}_j) f_{ij}({\rm cuts})\, .
\nonumber
\end{eqnarray} 

In the above equations we have introduced the following function:
\begin{eqnarray}
f_{ij}({\rm cuts}) &=& \int_{\rm cuts}d{\rm PS}~ d\sigma^{\rm pert}_{b\bar{b}} \otimes_b \left( D_{B_i} \otimes D^R_i \right)_b \nonumber\\
&&~~~~~ ~~~~~~~~~ ~~~~~~~ \otimes_{\bar{b}} \left(D_{B_j} \otimes D^R_j \right)_{\bar{b}} \, ,
\label{eq:fij}
\end{eqnarray} 
where $f_{ij}({\rm cuts}) = f_{ji}({\rm cuts})$ and $i,j=s,d$.

Finally, the result for $A_{\rm th}$ can be written as:
\begin{equation}
A_{\rm th} = {N^{++}({\rm cuts}) - N^{--}({\rm cuts}) \over N^{++}({\rm cuts}) + N^{--}({\rm cuts})} 
\equiv {A_{\rm num}\over A_{\rm den}} \, ,
\label{eq:Ath}
\end{equation}
where the function $A_{\rm num}$ reads:
\begin{eqnarray}
A_{\rm num} &=&  f_s^2 \Gamma_s^2f_{ss}({\rm cuts}) T(B_s\to B_s) T_s^- \nonumber\\
&+&   f_d^2 \Gamma_d^2f_{dd}({\rm cuts}) T(B_d\to B_d) T_d^- \nonumber\\
&+&   f_sf_d \Gamma_s\Gamma_d f_{sd}({\rm cuts}) \nonumber\\
&\times & \left[ T(B_d\to B_d) T_s^- +  T(B_s\to B_s) T_d^- \right]\, ,\label{eq:Anum}\\
A_{\rm den} &=& A_{\rm num}\left( T_q^- \to T_q^+ \right)~,~ q=s,d \, ,
\label{eq:Aden}
\end{eqnarray} 
and
\begin{equation}
T_q^\pm = T(\bar{B}_q \to B_q)  \pm T(B_q \to \bar{B}_q) \, .
\label{eq:T-}
\end{equation} 

Eqs.~(\ref{eq:Ath},\ref{eq:Anum},\ref{eq:Aden}) are our main result. In general, the three functions $f_{ss},f_{dd}$ and $f_{sd}$ are different from each other and therefore Eq.~(\ref{eq:Anum}) does not factorize. Clearly, the function $A_{\rm th}$ becomes dependent on the experimental cuts, too. In order to gain more inside into the properties of this result, we consider limiting cases first. In the two cases:
\begin{enumerate}
\item  $D_{B_s}=D_{B_d}$ and $D^R_s=D^R_d$ for any cuts,
\item  fully inclusive case (i.e. integration over the full phase space) for arbitrary fragmentation and decay functions,
\end{enumerate}
we have $f_{ss}=f_{dd}=f_{sd}$. In these two cases the functions $f_{ij}$ factor out completely from Eqs.~(\ref{eq:Anum},\ref{eq:Aden}) and cancel in Eq.~(\ref{eq:Ath}). Therefore, $A_{\rm th}$ becomes independent of the kinematics (i.e. cuts) and takes the usual form \cite{Grossman:2006ce}:
\begin{equation}
A_{\rm th} = {f_sT_s^- + f_dT_d^-  \over  f_sT_s^+ + f_dT_d^+  } ~~, ~~ {\rm in~cases}~1,2  \, .
\label{eq:Ath-usual}
\end{equation}

To get an insight into the general result Eqs.~(\ref{eq:Ath},\ref{eq:Anum},\ref{eq:Aden}), we consider the following simplification:
\begin{equation}
f_{ij}({\rm cuts}) \approx \phi_i({\rm cuts})\phi_j({\rm cuts}) \, ,
\label{eq:phi}
\end{equation}
i.e. the dependence on the two fragmentation channels factorizes. The motivation for this approximation follows the results of Ref.~\cite{Cacciari:2002pa}: in the case of single $B$ meson production in the same $p_T$ range, the partonic cross-section behaves as a power $d\sigma\sim 1/p_T^n$. In such a case the convolution becomes sensitive only to the $n$-th moment of the corresponding fragmentation function and the convolution becomes a simple product, i.e. $\phi_i \sim D_{B_i}^{(n)}$. In the following we take $n=5$. 

Within the approximation (\ref{eq:phi}), the same sign dimuon asymmetry reads:
\begin{equation}
A_{\rm th} = {f_s\phi_s({\rm cuts})T_s^- + f_d\phi_d({\rm cuts})T_d^-  \over  f_s\phi_s({\rm cuts})T_s^+ + f_d\phi_d({\rm cuts})T_d^+  } \, .
\label{eq:Ath-fin}
\end{equation}
The above equation has the virtue that it exhibits the essence of the general result (\ref{eq:Ath},\ref{eq:Anum},\ref{eq:Aden}) (i.e. explicit dependence on cuts and production stage effects) but in a simplified manner that allows direct interpretation. In particular, we see that all the dependence on cuts and $B$ meson production-stage effects can be absorbed into effective production fractions $f^{\rm eff}_i=f_i \phi_i({\rm cuts})$. It is for this reason we believe that our analysis underscores the need for a better understanding of the difference in the production fractions $f_i$ measured at the Tevatron and at the $Z$ pole.
\footnote{Please refer to sec. 3.1.3 of Ref.~\cite{TheHeavyFlavorAveragingGroup:2010qj} for a detailed overview of that difference and some of its implications.}

We next use the independence of (\ref{eq:Ath}) with respect to the normalization of the functions $\phi_i$ and divide the whole result by $\phi_d$. This way all $\phi$-dependence is absorbed in the ratio
\begin{equation}
{\phi_s\over \phi_d} = {D_{B_s}^{(n)} \over D_{B_d}^{(n)} } \equiv \kappa < 1 \, , 
\end{equation}
as follows from the discussion in sec.~\ref{sec:fragment}, and our default choice is $n=5$.

Expressing the time integrated amplitudes $T$ through the semileptonic asymmetries $A^q_{\rm SL}$ \cite{Grossman:2006ce}: $T_q^+ = Z_q/\Gamma_q^t$, $T_q^- = A^q_{\rm SL}Z_q/\Gamma_q^t$, with $Z_q \approx 1-1/(1+x_q^2)$, we get:
\begin{eqnarray}
A_{\rm th} &=& C_s A^s_{\rm SL} + C_d A^d_{\rm SL} \, ,\nonumber\\
C_s &=& {\kappa f_s Z_s \over \kappa f_s Z_s + f_d Z_d} \, ,\nonumber\\
C_d &=& {f_d Z_d \over \kappa f_s Z_s + f_d Z_d} \, .
\label{eq:Ath-final}
\end{eqnarray}

For consistency, the numerical values for the parameters $f_q({\rm Tev})$ and $x_q$ are as in Ref.~\cite{Abazov:2010hv}, while $f_i({\rm Z~pole})$ are taken from Ref.~\cite{TheHeavyFlavorAveragingGroup:2010qj}. We stress again that within the approximation (\ref{eq:phi}) all dependence on cuts and $B$ meson production effects is contained in the parameter $\kappa$.  
\begin{table}[htdp]
\caption{$A_{\rm th}$ as a function of the input parameters $f_i$ (Tevatron versus $Z$ pole) and $\kappa$. Only uncertainties from fragmentation are shown.}
\begin{center}
\begin{tabular}{|c|c|c|c|c|}
\hline\hline
$\kappa$ & 1.00 & 1.00 & 0.84 & 0.84 \\ 
\hline
$f_i$ measured at: & {\rm Tev} & {\rm $Z$ pole} & {\rm Tev} & {\rm $Z$ pole} \\
\hline
$A_{\rm th}\times 10^{-4} $ & -2.33 & -2.77 & $-2.54^{+0.10}_{-0.13}$ & $-2.98^{+0.10}_{-0.13}$ \\
\hline
$\left(A_{\rm th}-A^{(0)}_{\rm th}\right)/A^{(0)}_{\rm th} ~ [\%] $ & 0 & 19 & $9^{+6}_{-4}$ & $28^{+5}_{-5}$ \\
\hline
\end{tabular}
\end{center}
\label{table}
\end{table}

In table~\ref{table} we present $A_{\rm th}$ as a function of the parameters $f_i$ and $\kappa$. We present both its absolute value and the relative change with respect to the central value of $A^{(0)}_{\rm th}=(-2.33^{+0.5}_{-0.6})10^{-4}$ given in Ref.~\cite{Abazov:2010hv}. We have studied the sensitivity of our result due to variation of the fragmentation function parameters (i.e. $m_{d,s}$) and the moment $n$ of the fragmentation function. This is an approximately $\pm 5\%$ effect, as indicated in table \ref{table}.

For comparison, the measured value \cite{Abazov:2010hv} is $A_{\rm exp} = (-96 \pm 29) 10^{-4}$, where for simplicity, we have added the statistical and systematic errors in quadrature.

Our best prediction is $A_{\rm th} = (-3.0^{+0.6}_{-0.7}) 10^{-4}$. It includes $B$ meson fragmentation effects (i.e. $\kappa = 0.84$) and the values of the production fractions measured at the $Z$ pole
\footnote{Note also that the world average value for $f_d$ is very close to the one from the Z pole, while the one for $f_s$ is approximately the average of the two measurements \cite{TheHeavyFlavorAveragingGroup:2010qj}.}
\cite{TheHeavyFlavorAveragingGroup:2010qj}. For the sake of being more conservative, the uncertainty is derived by adding linearly the theoretical uncertainty quoted in Ref.~\cite{Abazov:2010hv} with the uncertainty from $B$ meson fragmentation quoted in table \ref{table}.

\section{Discussion}\label{sec:discussion}

It is obvious that the fragmentation effects analyzed in this paper cannot explain the large difference between the D\O\ measurement \cite{Abazov:2010hv} and the SM prediction. Moreover, additional sources of asymmetry are unlikely to be numerically significant (see appendix \ref{sec:app}).

The most attractive possibility for explaining this apparent discrepancy is bSM physics. Clearly, the results of this paper will directly propagate in any  analysis involving new physics, as long as the effects of bSM physics are confined to the mixing of the $B$ mesons, as expected. In the following we present the implication of the D\O\,  measurement - including the fragmentation effects studied in this paper - for the extraction of $A^s_{\rm SL}$. 

\begin{figure}[h]
  \centering
     \hspace{-1mm} 
   \includegraphics[width=3.5in]{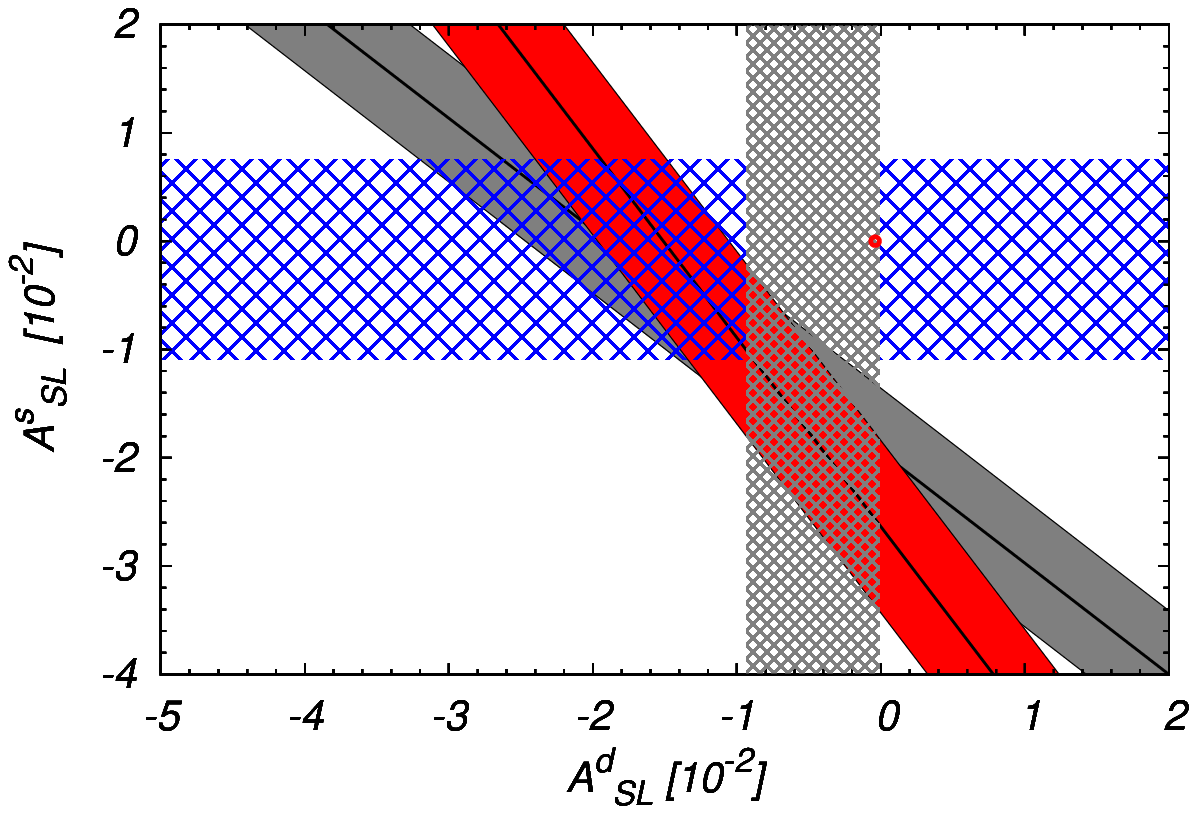} 
   \caption{The D\O\ measurement represented in the $A^d_{\rm SL} - A^s_{\rm SL}$ plane. The blue horizontal band is the constraint from the D\O\ measurement \cite{Abazov:2009wg} of $B_s^0\to \mu^+D_s^- X$; the light grey vertical band is the constraint from B-factories \cite{TheHeavyFlavorAveragingGroup:2010qj}. The solid black band is the one from Ref.~\cite{Abazov:2010hv} and the solid red band is our result. The small red circle is the SM value (uncertainty is  not to scale).}
   \label{figure}
%
%
     \hspace{-1mm} 
   \includegraphics[width=3.5in]{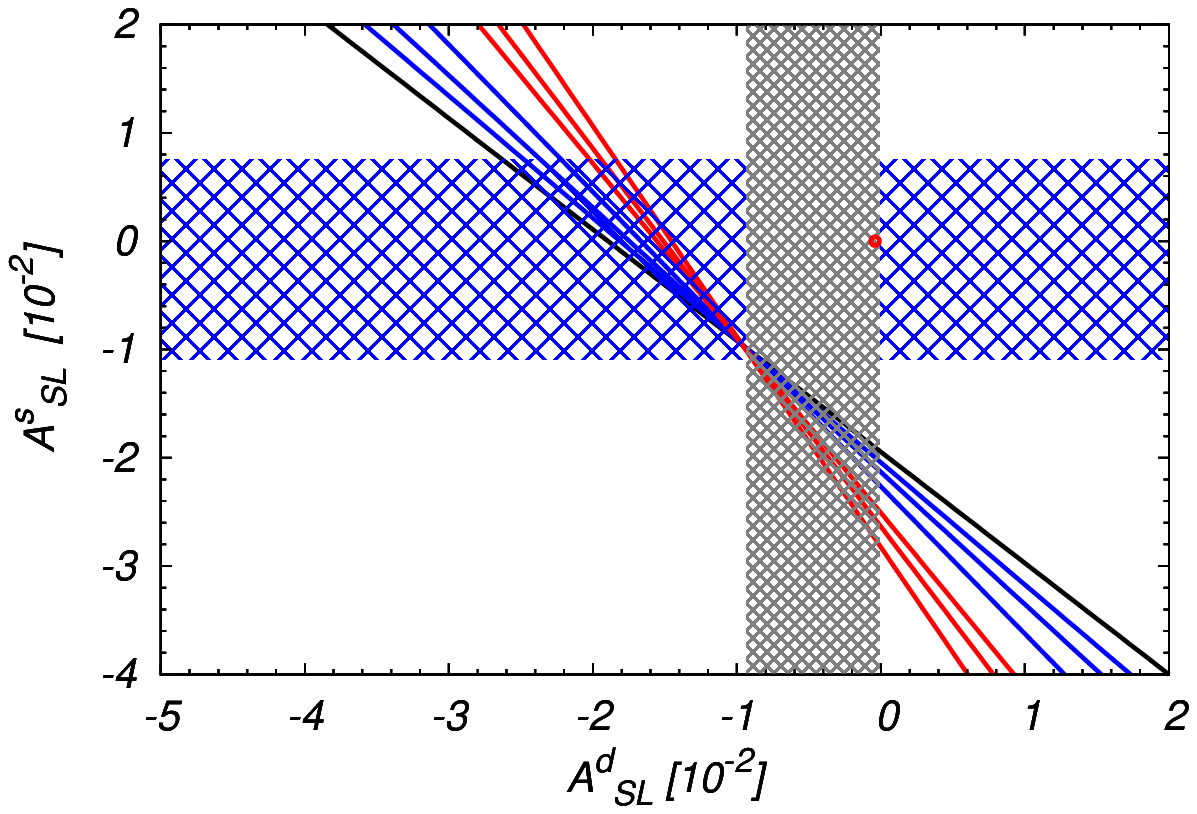} 
   \caption{Variation due to fragmentation (i.e. $\kappa$) and production fractions $f_i$. Plotted are the three scenarios described in the text. Central values are shown. The black line corresponds to the center of the dark grey band on figure~\ref{figure} and the central red line corresponds to the center of the red band on figure~\ref{figure}. }
   \label{figure-variation}
\end{figure}

On fig.~\ref{figure} we present the  D\O\ measurement in the $A^d_{\rm SL} - A^s_{\rm SL}$ plane. We give the existing constraints from independent measurements on $A^d_{\rm SL}$ (from B-factories \cite{TheHeavyFlavorAveragingGroup:2010qj}) with a light-grey band, and on $A^s_{\rm SL}$  (from the D\O\ measurement \cite{Abazov:2009wg} of $B_s^0\to \mu^+D_s^- X$) with a blue band. We plot the D\O\ dimuon asymmetry measurement in two ways. First we do not include fragmentation effects, i.e. $\kappa=1$, and take production fractions as measured at the Tevatron (with a dark-grey band); this is the same as figure 17 in Ref.~\cite{Abazov:2010hv}. Second, we plot the measurement including fragmentation effects, i.e. $\kappa = 0.84$, and with production fractions as measured at the Z pole (with a red band).

In all cases the plots are obtained by solving Eq.~(\ref{eq:Ath-final}) for $A^s_{\rm SL}$ and replacing $A_{\rm th}$ with the measured value $A_{\rm exp} = (-96 \pm 29) 10^{-4}$. In units of $[10^{-2}]$ we get:
\begin{equation}
A^s_{\rm SL} = \alpha A^d_{\rm SL} + \beta \, .
\label{eq:plot}
\end{equation}
The coefficients $\alpha$ and $\beta$ are given in table \ref{table:plot}. The width of the two bands reflects the uncertainty in the measurement only. The end-lines of the bands on figure ~\ref{figure} can be obtained by adding $\pm 0.796$ (for the solid red band) and $\pm 0.588$ (for the solid black band) to the coefficient $\beta$ of the corresponding central line.
\begin{table}[htdp]
\caption{The coefficients in Eq.~(\ref{eq:plot}). {\it This paper} is the center of the solid red band on figure ~\ref{figure} and the central red line on figure ~\ref{figure-variation}. {\it D\O} (same as figure 17 in Ref.~\cite{Abazov:2010hv}) is the center of the solid black band on figure ~\ref{figure} and the black line on figure ~\ref{figure-variation}. {\it Blue C,1,2} are the blue central and end-lines on figure ~\ref{figure-variation}. {\it Red 1,2} are the two red end-lines on figure ~\ref{figure-variation}.}
\begin{center}
\begin{tabular}{|c|c|c|c|c|c|c|c|}
\hline\hline
\rm{Line} & \it{This paper} & \it{D\O} & \it{Blue C} & \it{Blue 1} & \it{Blue 2} & \it{Red 1} &\it{Red 2}  \\ 
\hline
$\alpha $ & -1.745 & -1.027 & -1.221 & -1.360 & -1.128 & -1.944 & -1.613 \\
\hline
$\beta   $ & -2.636 & -1.946 & -2.132 & -2.266 & -2.043 & -2.826 & -2.509  \\
\hline
\end{tabular}
\end{center}
\label{table:plot}
\end{table}

To get an insight into the sensitivity separately to fragmentation (through $\kappa$) and production fractions (through the values of $f_q$), on figure ~\ref{figure-variation} we plot the central values for the following scenarios:
\begin{enumerate}
\item (black line): $\kappa=1$ and production fractions $f_i({\rm Tev})$, as in Ref.~\cite{Abazov:2010hv}\, ,
\item (with blue lines): $f_i({\rm Tev})$ and $\kappa=0.84^{+8\%}_{-10\%}$ \, ,
\item (with red lines): $f_i({\rm Z\, pole})$ and $\kappa=0.84^{+8\%}_{-10\%}$\, .
\end{enumerate}

The uncertainty on $\kappa$ corresponds to the $\pm 5\%$ uncertainty on $A_{\rm th}$ specified in table \ref{table}. We do not present the uncertainties from the measurement; they can be inferred from fig.~\ref{figure}.

There are two important features that arise from our analysis. First, there is a point where the sensitivity to fragmentation functions/production fractions is zero. Strikingly, this point is precisely at the place where the three independent measurements agree with each other. An important implication is that the analysis in this region is {\it robust} with respect to these parameters.

Second, we note that each one of the two effects studied in this paper: $B$ fragmentation and sensitivity to the values of the B meson production fractions, pushes the band of the dimuon asymmetry measurement further away from the SM expectation in the region of small $\vert A^d_{\rm SL}\vert $. This result further strengthens the arguments given in Ref.~\cite{Dobrescu:2010rh} that new physics contributions are unlikely to only affect $A^s_{\rm SL}$, as one might naively expect; see also Refs.~\cite{Bevan:2010gi} and \cite{Lunghi:2010gv}. Indeed, from figures~\ref{figure},\ref{figure-variation} we conclude that the effects considered in this paper make it even more likely that the values of {\it both} $A^d_{\rm SL}$ and $A^s_{\rm SL}$ are significantly away from the SM expectations.

\vskip 5mm
{\bf Note added.} After this article was submitted for publication, a new measurement of the ratio $f_s/f_d$ was performed by the LHCb collaboration \cite{Aaij:2011hi}. The result of Ref.~\cite{Aaij:2011hi} is in complete agreement with the measurements of $f_{s,d}$ at the Z pole. This new measurement confirms our QCD factorization--based arguments in favor of the values of the fragmentation fractions as measured at the Z pole. Regarding figure~\ref{figure-variation}, the measurement of Ref.~\cite{Aaij:2011hi} favors the set of red over blue lines.

\begin{acknowledgments}
The author thanks the Yang Institute for Theoretical Physics at Stony Brook University where this work was initiated. He also would like to thank Dmitri Tsybychev for a number of clarifications regarding Ref.~\cite{Abazov:2010hv}, L.~Dixon for useful suggestions and M.~Cacciari, Y.~Grossman, K.~Melnikov and U.~Nierste for discussions.
\end{acknowledgments}

\appendix
\section{Backgrounds}\label{sec:app}

Consider same-sign dimuon final states that contain $\mu^+\mu^+$ or $\mu^-\mu^-$ pairs with equal probability. Such events are backgrounds and they only contribute to the denominator of Eq.~(\ref{eq:Ab-sl}). In presence of such backgrounds the expression for the asymmetry can be cast in the form:
\begin{eqnarray}
&& A_{\rm th} = A_{\rm th}^{({\rm bkg}=0)} \times {1\over 1+ \delta({\rm cuts})}  \, , \nonumber\\
&& \delta({\rm cuts}) = {N^{++}_{\rm bkg}+N^{--}_{\rm bkg} \over N^{++}_{\rm signal}+N^{--}_{\rm signal} } \approx {N^{++}_{\rm bkg} \over N^{++}_{\rm signal} }\, ,
\label{eq:background}
\end{eqnarray}
where $A_{\rm th}^{({\rm bkg}=0)}$ is the background-free asymmetry studied in the previous sections. The correction $\delta$ is estimated in Ref.~\cite{Abazov:2010hv} to be $\sim {\cal O}(10\%)$. From Eq.~(\ref{eq:background}) it is evident that the effect of backgrounds is to decrease the value of $A_{\rm th}$. The background correction is also cut dependent. 

Higher order effects can be an additional source of backgrounds. An example are partonic final states of the type $bb+X$ or $\bar{b}\bar{b}+X$ followed by two ``right" decays. Such contributions are likely small, contributing at most few percent to $A_{\rm th}$ and likely much less. The reason their contribution is small is that such states are generated in processes suppressed by two powers of the strong coupling $\alpha_S$, for example in the leading order contribution to the process $p\bar{p} \to b\bar{b}b\bar{b}$. Alternatively, one can think of such processes as ``usual" $b\bar{b}$ symmetric final states where, for example, the $b$ quark perturbatively fragments into a $\bar{b}$ \cite{Melnikov:2004bm} thus producing an inclusive final state with two $\bar{b}$ quarks. 

On the other side, the ``gain" factor due to the absence of ``wrong" type of decays is small for $B_s$ and $B_d$ mesons. It can be estimated as follows. The suppression factor of a ``wrong'' decay with respect to a ``right'' decay is given by the ratio of time integrated probabilities (see \cite{CPbook,Grossman:2006ce}):
\begin{eqnarray}
C_q \equiv { T[B_q\to \bar{B_q}]\over T[B_q\to B_q]} \approx {x_q^2\over 2+x_q^2} \, ,
\end{eqnarray} 
for $q=s,d$. Numerically $C_s\approx 1$ and $C_d\approx 0.2$.

\end{document}